\documentclass[a4paper,twocolumn,showpacs,superscriptaddress,floatfix]{revtex4-1}
\usepackage{graphicx}
\usepackage{epsf}
\usepackage{epsfig}
\usepackage{amssymb,amsmath}
\usepackage[usenames]{color}
\usepackage{amssymb}
\usepackage{times}

\newcommand{\be}{\begin{equation}}
\newcommand{\ee}{\end{equation}}
\newcommand{\bea}{\begin{eqnarray}}
\newcommand{\eea}{\end{eqnarray}}
\newcommand{\nn}{\nonumber}

\newcommand{\Sp}{{\mathcal S}}

\font\tenscr=rsfs10 scaled1100
\font\sevenscr=rsfs7 
\font\fivescr=rsfs5 
\skewchar\tenscr='177
\skewchar\sevenscr='177
\skewchar\fivescr='177
\newfam\scrfam
\textfont\scrfam=\tenscr
\scriptfont\scrfam=\sevenscr
\scriptscriptfont\scrfam=\fivescr

\begin{document}

\title{
A Young-Laplace law for black hole horizons
}

\author{Jos\'e Luis Jaramillo}
\affiliation{
Max-Planck-Institut f{\"u}r Gravitationsphysik, Albert Einstein
Institut, Am M\"uhlenberg 1 D-14476 Potsdam Germany 
}

\begin{abstract}
Black hole horizon sections, modelled as marginally
outer trapped surfaces (MOTS), possess a notion of stability
admitting a spectral characterization.
Specifically, the ``principal eigenvalue'' $\lambda_o$ 
of the MOTS-stability operator (an elliptic operator on 
horizon sections) must be non-negative.
We discuss the expression of $\lambda_o$ for axisymmetric stationary black hole horizons 
and show that, remarkably, it presents the form of the 
Young-Laplace law for soap bubbles in equilibrium, 
if $\lambda_o$ is identified with a formal pressure difference between the
inner and outer sides of the ``bubble''. In this view, that endorses the 
existing fluid analogies for black hole horizons, MOTS-stability is 
interpreted as a consequence of a pressure 
increase in the black hole trapped region.

\end{abstract}


\maketitle

\section{Introduction}
Mechanical fluid analogies have played an
important role in building our intuition
of black hole (BH) horizon dynamics. 
The comparison with a rotating liquid drop was early discussed
\cite{Smarr:1972kt}, providing 
an interpretation of 
the BH surface gravity as the corresponding liquid
surface tension. 
More systematically, the analogy of the BH horizon with a
2-dimensional viscous fluid
was developed in \cite{Damou, Price86,ThoPriMac86}
(and references therein) 
building the so-called ``membrane paradigm'', of particular
interest in astrophysical BH dynamics.
Remarkably, aspects of the latter ``membrane perspective''
have been recently revisited in higher dimensional settings 
in the context of the CFT/AdS duality, namely
the correspondence between the (bulk) gravitational description of 
an asymptotically Anti-de Sitter spacetime and the dynamics
of an appropriate conformal field theory at its boundary (e.g. \cite{Huben11}). 
Related analogies of BH horizons as ``soap bubbles'' 
can be found in \cite{Eard98,Eardley:2002re} and, particularly
interesting in our present context, have led 
to the discussion of the Gregory-Laflamme 
instability of black strings in terms of the classical
fluid Rayleigh-Plateau instability \cite{CarDia06}.

Here we further support these
analogies by interpreting the stability of
stationary
BHs in terms of the Young-Laplace
law for ``soap bubbles''. This relates 
the pressure difference at the interface between fluids in equilibrium
to the interface shape 
\bea
\label{e:Young-Laplace}
\Delta p =  p_{\mathrm{inn}}-p_{\mathrm{out}} = \gamma \left(1/R_1 + 1/R_2  \right) \ ,
\eea 
where at any interface point 
$\Delta p$ is the difference between the inner and outer
pressures ($p_{\mathrm{inn}}$ and $p_{\mathrm{out}}$),
$\gamma$ is the surface tension and $R_{i=1,2}$ 
are the principal curvature
radii (with normal vector pointing
outwards).
Specifically, we show that MOTS-stability \cite{AndMarSim}
of stationary BH horizons,  
characterised by the non-negativity
of the so-called principal eigenvalue
$\lambda_o$ of the MOTS-stability operator $L_{\cal S}$ (see below), can 
be discussed in terms of the Young-Laplace
law in Eq. (\ref{e:Young-Laplace}) if $\lambda_o$
is identified with a formal pressure difference $\Delta p$. 
This provides a first step in the systematic spectral analysis 
of the MOTS-stability operator, as well as a suggestive interpretation
shift that casts this geometric stability problem on fluid physical grounds.

\section{MOTS, stability and quasi-local horizons} 
Let us introduce the specific notion of stability
here discussed.
Let us consider a $d-$dimensional spacetime $({\cal M}, g_{ab})$
with Levi-Civita connection $\nabla_a$ and a closed spacelike 
$(d-2)$-surface ${\cal S}$ (we make $G=c=1$). Let $q_{ab}$ denote 
the induced metric on ${\cal S}$, and $D_a$ and $R$ its associated 
Levi-Civita connection and Ricci scalar.
We span the normal plane $T^\perp{\cal S}$ by (future) outgoing $\ell^a$ 
and ingoing $k^a$ null
vectors, normalised as $\ell^a k_a =-1$. 
Expansions in the
outgoing and ingoing directions are 
$\theta^{(\ell)}=q^{ab}\nabla_a \ell_b$ and $\theta^{(k)}=q^{ab}\nabla_a k_b$.

The surface ${\cal S}$ is called (strictly)  outer trapped iff $\theta^{(\ell)} < 0$ and 
a marginally outer trapped surface (MOTS)
iff $\theta^{(\ell)}=0$. 
MOTSs possess a natural notion of stability \cite{AndMarSim}:
a MOTS surface ${\cal S}$ is said to be (strictly) stable  if it admits a 
deformation along $k^a$ that is outer trapped or, equivalently, 
a deformation along $-k^a$ that is fully non-trapped.
In other words, the MOTS ${\cal S}$ is stable if there exists a positive 
function $\psi$ on ${\cal S}$ such that
$\delta_{\psi (-k)} \theta^{(\ell)} > 0$, where $\delta$ denotes the deformation 
operator of the
surface ${\cal S}$ discussed in \cite{AndMarSim,BooFai07}. This
notion of stability admits a spectral characterization in terms of the
MOTS-stability operator $L_{\cal S}$ defined on ${\cal S}$ as
\bea
\label{e:MOTS-stability_operator}
L_{\cal S} \psi &\equiv& \delta_{\psi (-k)} \theta^{(\ell)} 
= \left[- D^aD_a  + 2 \Omega^{(\ell)}_a D^a \right. \\
&&-\left.\left(\Omega^{(\ell)}_a  {\Omega^{(\ell)}}^a -D^a  \Omega^{(\ell)}_a
-\frac{1}{2}{}R + G_{ab}\ell^ak^b\right)\right] \psi \nn \ ,
\eea
where 
$\Omega^{(\ell)}_a = -k^c {q^b}_a\nabla_b \ell_c $ is
the connection in  $T^\perp{\cal S}$ \cite{BooFai07}
and $G_{ab}$ is the Einstein tensor.
The eigenvalues are generically complex numbers ($L_{\cal S}$ is 
not self-adjoint). However the {\em principal eigenvalue} $\lambda_o$, 
namely the eigenvalue with smallest real part, can be shown 
to be real \cite{AndMarSim}. MOTS-stability of
${\cal S}$ is then characterised by the non-negativity of $\lambda_o$
\cite{AndMarSim}
\bea
\label{e:spectral_MOTS_stability}
\lambda_o \geq 0 \ \ ,
\eea
with positive principal eigenfunction $\phi_o$ 
(i.e. $L_{\cal S} \phi_o = \lambda_o \phi_o$). 
We also define an 
operator $L^*_{\cal S}$ obtained from $L_{\cal S}$ by imposing Einstein equations, 
$G_{ab} + \Lambda g_{ab} = 8\pi T_{ab}$, but dropping the (explicit) 
presence of the cosmological constant $\Lambda$:
\bea
\label{e:MOTS-stability_operator_*}
L^*_{\cal S} \psi &\equiv& 
\left[- D^aD_a  + 2 \Omega^{(\ell)}_a D^a \right. \\
&&-\left.\left(\Omega^{(\ell)}_a  {\Omega^{(\ell)}}^a -D^a  \Omega^{(\ell)}_a
-\frac{1}{2}{}R + 8\pi T_{ab}\ell^ak^b\right)\right] \psi \nn \ .
\eea

Quasi-local models for BHs~\cite{Hay94,AshKri04} can be constructed by considering
marginally trapped tubes (MTT), namely 
hypersurfaces ${\cal H}$ admitting a foliation
$\{{\cal S}_t\}$ by closed MOTS, i.e.
${\cal H} = \bigcup_{t\in\mathbb{R}} {\cal S}_t$. 
Under the null energy condition and assuming
MOTS stability (the {\em outer condition} in \cite{Hay94}), 
MTTs are either null or spacelike hypersurfaces
\cite{Hay94,BooFai07}. 
The former corresponds to non-expanding horizons
whereas the latter, 
under the future condition $\theta^{(k)}\leq 0$,
are dynamical expanding ones.
We focus here on the equilibrium case,
where the null horizon ${\cal H}$ is generated
by the null vector $\ell^a$ and the intrinsic geometry remains 
invariant under $\ell^a$: ${\cal L}_\ell q_{ab}=0$. Crucially for our discussion,
any foliation of ${\cal H}$ defines a foliation by MOTS.
This freedom will be exploited in Theorem 1 below.  
We introduce the {\em surface gravity} $\kappa^{(\ell)}$ as the non-affinity
coefficient of $\ell^a$, i.e. $\ell^b \nabla_b \ell^a = \kappa^{(\ell)}\ell^a$,
with $\kappa^{(\ell)} = -k^a\ell^b \nabla_b \ell_a$. 

We will consider a stronger notion of stationarity than that of 
non-expanding horizons, by requiring also the
extrinsic geometry of the null ${\cal H}$ to be invariant under a certain
$\ell^a$ fixed up to a constant rescaling.
This defines an {\em isolated horizon} (IH) \cite{AshKri04,AshBeeLew02}.
More specifically, we require the invariance of the unique connection $\hat{\nabla}_a$
induced on the non-expanding horizon ${\cal H}$ by the ambient one $\nabla_a$:
$[{\cal L}_\ell,\hat{\nabla}_a]=0$. This implies the invariance
of $\Omega^{(\ell)}_a$ and $\kappa^{(\ell)}$, i.e. ${\cal L}_\ell\Omega^{(\ell)}_a
= {\cal L}_\ell\kappa^{(\ell)}=0$, and the angular
constancy of $\kappa^{(\ell)}$: $D_a\kappa^{(\ell)}=0$. 
IHs constitute the model for
stationary BH horizons discussed here. This includes
in particular Killing horizons, in which $\ell^a$ can be extended to a symmetry
in the spacetime neighbourhood of ${\cal H}$.

\section{$\lambda_o$ eigenvalue for axisymmetric IHs}
The sign of the principal eigenvalue $\lambda_o$ 
controls  MOTS-stability, as expressed in 
(\ref{e:spectral_MOTS_stability}). 
It is therefore of interest to have an explicit 
expression of $\lambda_o$ in terms
of the geometry of ${\cal S}$. Although this is a challenging 
problem when considered in full generality, the
very important case of stationary and axisymmetric BH horizons
is addressed by the following result~\cite{Note1}:

\medskip

{\bf Theorem 1} (Reiris \cite{Reiris13}).
{\em Given an axisymmetric IH ${\cal H}$ with null generator $\ell^a$
and non-affinity coefficient $\kappa^{(\ell)}$:}
\begin{itemize}
{\em \item[i)] There exists an (axisymmetric) foliation 
${\cal H}=\bigcup_{t} {\Sp}^o_t$ by MOTSs ${\Sp}^o_t$ with constant
ingoing expansion $\theta^{(k)}$.}
{\em \item[ii)] The principal eigenvalue $\lambda_o$ evaluated on 
sections ${\Sp}^o_t$ is
\bea
\label{e:lambda_o}
\lambda_o = - \kappa^{(\ell)} \theta^{(k)} \ .
\eea
}
{\em \item[iii)] The principal eigenfunction $\phi_o$ is given by
$\phi_o = e^{2 \chi}$, 
with $\Omega_a^{(\ell)} =  z_a + D_a \chi$ on ${\Sp}^o_t$ , where $D^az_a=0$.}
\end{itemize}
The result holds in any dimensions, under the topological
condition in \cite{Mars:2012sb} of ${\cal H}$ being 
foliated by closed MOTSs.
Note that $\lambda_o$ does not depend on the section of ${\cal H}$
\cite{Mars:2012sb}, though the particular form (\ref{e:lambda_o}) 
only holds in the preferred
foliation $\{{\Sp}^o_t\}$ in Theorem 1.

\section{Young-Laplace law for stationary horizons}

\subsection{BH surface tension and mean curvature}
Let us first rewrite expression (\ref{e:lambda_o}) in the 
following way
\bea
\label{e:lambda_o_YL}
\lambda_o/(8\pi) = \kappa^{(\ell)}\!/(8 \pi) \;(- \theta^{(k)}) \ .
\eea
The right-hand-side presents then a particularly 
suggestive form  when compared with the Young-Laplace law in 
(\ref{e:Young-Laplace}). 
First, from the first law of BH
thermodynamics, namely $\delta M = \kappa^{(\ell)}\!/(8 \pi) \delta A + \Omega \delta J$,
the factor $\kappa^{(\ell)}\!/(8 \pi)$
is identified in \cite{Smarr:1972kt} as an effective BH surface tension
\bea
\label{e:BHsurfacetension}
\gamma_{_{\mathrm{BH}}} = \kappa^{(\ell)}\!/(8 \pi) \ ,
\eea
using its standard equivalence with an energy surface density.
Such thermodynamical identification is consistent
with the purely mechanical view provided by 
the analogy of BH horizons as $2-$dimensional viscous 
fluids in the membrane paradigm 
\cite{Damou, Price86,ThoPriMac86}.
In the latter, the understanding of the evolution equations for 
$\theta^{(\ell)}$ and $\Omega^{(\ell)}_a$ as, respectively, energy and momentum
(Damour-Navier-Stokes) balance equations requires 
the interpretation of $\kappa^{(\ell)}\!/(8 \pi)$ as a pressure of the 
$2-$dimensional fluid, i.e. a mechanical surface tension.
 
Second, regarding the second factor in  (\ref{e:lambda_o_YL}),
let us consider the section ${\cal S}^o_t$ 
provided by point {\em i)} in Theorem 1, and let us extend it to a $(d-1)$-dimensional 
spatial slice $\Sigma_t$ in the bulk. Such 
$\Sigma_t$ can be locally boosted so that
the IH null generator $\ell^a$ and the 
ingoing null normal 
to ${\cal S}^o_t$ are respectively written as $\ell^a = n^a + s^a$ 
and $k^a = (n^a - s^a)/2$,
with $n^a$ the timelike
normal to $\Sigma_t$ and $s^a$ the outgoing spacelike normal to ${\cal S}^o_t$
tangent to $\Sigma_t$. The mean curvature $H$
of $({\cal S}^o_t, q_{ab})$ into $(\Sigma_t, \gamma_{ab})$, with $\gamma_{ab}$
induced from the ambient $g_{ab}$, is written as
\bea
H =  q^{ab}\nabla_a s_b =  \tilde{D}_a s^a \ ,
\eea
with $\tilde{D}_a$ the connection compatible with $\gamma_{ab}$.
For a $2-$surface embedded in an Euclidean 3-space, the form
$H=(1/R_1+1/R_2)$ in (\ref{e:Young-Laplace}) is recovered.
Combining $\theta^{(\ell)}$ and $\theta^{(k)}$, we write
$H = - \theta^{(k)} +  \frac{1}{2} \theta^{(\ell)}$, so that in our 
MOTS $\theta^{(\ell)}=0$ case
\bea
\label{e:ExtrinsicCurvarture}
H = - \theta^{(k)} \ .
\eea
From (\ref{e:BHsurfacetension}) and (\ref{e:ExtrinsicCurvarture})
we see that (\ref{e:lambda_o_YL}) matches the form (\ref{e:Young-Laplace}) 
of the Young-Laplace law, if $\lambda_o/(8\pi) $ is {\em formally} identified 
with a pressure difference between the {\em interior} and the {\em exterior}
of the BH horizon.
We justify now such a heuristic identification.

\subsection{The principal eigenvalue $\lambda_o$ as a pressure}
The principal eigenvalue $\lambda_o$ admits the interpretation of a pressure.
First we note that $\lambda_o$ shares physical
nature with the cosmological constant $\Lambda$. 
Indeed, the (explicit) effect of switching-on 
the cosmological constant, as compared with the reference 
situation in absence of $\Lambda$, is to produce a shift in the eigenvalue
$\lambda_o$ characterising MOTS-stability~\cite{Note2}
\bea
\label{e:SpectralShift}
L_{\cal S} \phi = \lambda \phi \ \ , \ \ 
L^*_{\cal S} \phi = \lambda^* \phi \ \ \Longrightarrow \ \ 
\lambda_o^* =  \lambda_o + \Lambda \ ,
\eea
that follows from (\ref{e:MOTS-stability_operator}) and (\ref{e:MOTS-stability_operator_*})
when imposing $G_{ab} +\Lambda g_{ab} = 8\pi T_{ab}$.
Therefore, physical dimensions of 
$\Lambda$ are shared by $\lambda_o$.

Second, the cosmological constant $\Lambda$ admits the natural
interpretation of a pressure, $p_{\mathrm{cosm}} = -\Lambda/(8\pi)$,
for a perfect fluid stress-energy tensor. Based on these remarks,
we propose the interpretation of $\lambda_o/(8\pi)$ as a pressure,
specifically a pressure difference  between the
interior and exterior of the BH horizon
\bea
\label{e:pressure_diff}
\Delta p = p_{\mathrm{inn}}-p_{\mathrm{out}} \equiv \lambda_o/(8\pi) \ ,
\eea
with  $p_{\mathrm{inn}}$  and $p_{\mathrm{out}}$ the {\em formal} inner and outer pressures.

\subsection{The MOTS-stability operator $L_{\cal S}$ as a ``Pressure Operator''}
Beyond the interpretation of $\lambda_o$ in (\ref{e:pressure_diff}),
the whole stability operator $L_{\cal S}$ can be understood as
a ``pressure operator''.
To justify this claim, we consider the equation of
a MTT. The horizon evolution vector $h^a$, tangent to 
${\cal H}= \bigcup_{t\in\mathbb{R}} {\cal S}_t$
and normal to MOTS sections ${\cal S}_t$,  Lie-drags the section ${\cal S}_t$ to
${\cal S}_{t+\delta t}$. It can be written as $h^a = \ell^a - Ck^a$, where $C$ is a 
(dimensionless) function on ${\cal H}$ such that the MTT is null, spacelike or
timelike for $C=0$, $C>0$ or $C<0$, respectively.
The MTT condition $\delta_h \theta^{(\ell)}=0$ is then expressed 
in terms of the MOTS-stability operator $L_{\cal S}$. By using
$\delta_h \theta^{(\ell)} =  \delta_\ell \theta^{(\ell)} -\delta_{C k} \theta^{(\ell)}$,
the MTT condition is rewritten as $\delta_{C(-k)} \theta^{(\ell)} = - \delta_\ell \theta^{(\ell)}$, 
so that
\bea
\label{e:MTT_condition}
L_{\cal S} C = {\sigma^{(\ell)}}_{ab}{\sigma^{(\ell)}}^{ab} + 8\pi T_{ab}\ell^a\ell^b \ \ ,
\eea
where ${\sigma^{(\ell)}}_{ab}={q^c}_a {q^d}_b \nabla_c \ell_d -1/(d-2) \theta^{(\ell)}q_{ab}$
is the shear associated with the outgoing null normal and we have
made use of the null Raychaudhuri equation.
The right-hand-side of Eq. (\ref{e:MTT_condition}) 
fixes the physical dimensions of the stability operator as
 $[L_{\cal S}/(8\pi)]=\mathrm{Energy}\cdot\mathrm{Time}^{-1}\cdot\mathrm{Area}^{-1}$,
($G=c=1$). 
Such an interpretation is natural in dynamical scenarios, where
the horizon growth is controlled by the presence of matter or 
gravitational energy fluxes.
In purely stationary contexts, as in the spectral problem of (\ref{e:SpectralShift}),
physical dimensions of $L_{\cal S}$ can be recast in a better suited form 
by simply noting $\mathrm{Energy}\cdot\mathrm{Time}^{-1}\cdot\mathrm{Area}^{-1}
\approx \mathrm{Force}\cdot \mathrm{Area}^{-1}$, so 
that
\bea
\label{e:physical_dimensions_L_S}
[L_{\cal S}/(8\pi)]=\mathrm{Pressure} \ .
\eea
This provides additional support
to the proposed physical interpretation of (the real) $\lambda_o/(8\pi)$ as a pressure.
But, in addition, it also suggests a role of the whole spectrum of $L_{\cal S}$
(including complex eigenvalues) in horizon stability issues.

\subsection{MOTS-stability from a BH Young-Laplace law perspective}
We can now revisit MOTS-stability for stationary axisymmetric  
BHs in the following soap-bubble analogy form: 

\medskip

{\bf BH Young-Laplace ``law''}: {\em
For stationary axisymmetric IHs, there exists 
a foliation in which the identifications
\bea
\label{e:analogy}
\!\!\!\!\!\!\!\!\frac{\kappa^{(\ell)}}{8\pi} \to \gamma_{_{\mathrm{BH}}} \ , \ -\theta^{(k)}\to H 
\ , \frac{\lambda_o}{8\pi}\to \Delta p = p_{\mathrm{inn}}-p_{\mathrm{out}}\ ,
\eea 
permit to recast the principal eigenvalue in the form 
of a Young-Laplace law:
$\Delta p = p_{\mathrm{inn}}-p_{\mathrm{out}} = \gamma_{_{\mathrm{BH}}} H$.
In this view,  MOTS-stability ($\lambda_o\geq 0$)
is  interpreted as the result of an increase in the pressure  
of the BH trapped region.
}

\section{Perspectives from a Young-Laplace view}
Apart from the appeal of casting Theorem 1 in the physical terms of 
equilibrium bubbles, the main outcome of the 
Young-Laplace perspective is the identification of 
$\lambda_o$ as a pressure.
This interpretation extends beyond 
stationarity and axisymmetry, providing a new twist 
on MOTS-stability that suggests 
new avenues and questions motivated by the fluid analogy. 
The heuristic proposals in the  rest of the article illustrate this.

\subsection{BH horizon dynamical timescale}
The identification of $\kappa^{(\ell)}\!/(8 \pi)$ as a surface 
tension, together with the integrated expressions for
the BH mass $M = 2\kappa^{(\ell)}\!/(8 \pi) A + 2\Omega J$, 
led Smarr~\cite{Smarr:1972kt} to consider BH horizon
instabilities in analogy with the case of rotating
liquid drops.

Although MOTS-stability does not correspond to the notion of dynamical
stability, it provides a condition for equilibrium that can be used to
estimate the characteristic timescale of dynamical perturbations. 
This is illustrated for fluids in the Rayleigh-Plateau
instability, where the timescale $\tau_{_{\mathrm{RP}}}$ of the zero-mode dominating
at large times can be determined solely from the equilibrium 
Young-Laplace law: $\tau_{_{\mathrm{RP}}} = \sqrt{4\pi a^3 \rho/\gamma}$, with
$a$ the radius of the fluid jet, $\rho$ its density and $\gamma$
the surface tension. 
In this spirit, our geometrical
setting suggests the following proposal for a BH horizon 
dynamical timescale
\bea
\label{e:dyn_timescale}
\tau_{\mathrm{dyn}} \equiv \sqrt{1/\Delta p}=\sqrt{8\pi/\lambda_o} \ \ .
\eea
If this timescale corresponds to an instability, or rather to
a relaxation process, must be determined by other methods (e.g. \cite{Hollands:2012sf}).
The first case is illustrated
by the Gregory-Laflamme
instability of $d$-dimensional black strings, where
$\lambda_o= R_{S^{d-3}(r_{_{\mathrm{H}}})}/2 = (d-3)(d-4)/(2 r_{_{\mathrm{H}}}^2)$, 
with  $r_{_{\mathrm{H}}}$ the horizon areal radius, and (\ref{e:dyn_timescale}) produces
$\tau_{_{\mathrm{BS}}}=\sqrt{16\pi/[(d-3)(d-4)]}\;r_{_{\mathrm{H}}}$.
For $d=5$ 
\bea
\tau_{_{\mathrm{BS}}}=\sqrt{8\pi}\;r_{\mathrm{H}}= \sqrt{8\pi \cdot 4M^2} =
\sqrt{M/\gamma_{_{\mathrm{BH}}}} \ ,
\eea
where the last expression stresses the analogy with the Rayleigh-Plateau 
instability shown in \cite{CarDia06}, when introducing the effective mass
$m_{\mathrm{eff}}= 4\pi a^3 \rho$ in $\tau_{_{\mathrm{RP}}}$ above.
Regarding the stable case,
Reissner-Nordstr\"om ($d=4$) provides a non-trivial example
in which (\ref{e:dyn_timescale}) leads to a dynamical timescale 
\bea
\label{e:tau_RN}
\tau_{_{\mathrm{RN}}}=
\frac{\sqrt{4\pi}\;\left(M+\sqrt{M^2-Q^2}\right)^2}
{\sqrt{M(M+\sqrt{M^2-Q^2}) -Q^2}} \ .
\eea
The shortest timescale occurs at $Q/M=\sqrt{3}/2$. Interestingly, this number
coincides with the value for heat capacity change of sign
in Reissner-Nordstr\"om \cite{Davies:1978mf}
(and with Smarr's proposal for the critical
$J/M^2$ for  Kerr ``rotating instabilities'').

\subsection{Full spectral analysis of $L_{\cal S}$ and BH horizon instabilities}
Beyond the role of $\lambda_o$ in setting a dominating timescale, 
the full spectrum of $L_{\cal S}$ may provide a more refined probe
into the stability/dynamical properties of the horizon. This is suggested by the
presentation of the whole stability operator in (\ref{e:physical_dimensions_L_S})
as a ``pressure operator''. 
Such an approach is particularly rich in the rotating case since
higher eigenvalues $\lambda_{n>o}$'s are then generically complex,
due to the  $2 \Omega^{(\ell)}_a D^a$ term, 
with imaginary part encoding rotational information~\cite{Note3}.
In particular, it is of interest 
to study a possible imprint 
of superradiance in the imaginary
part of the spectrum. 
In brief, we propose here the systematic 
study of the full spectrum of  $L_{\cal S}$ in a line of research
that, inspired by the inverse spectral problem for the Laplacian 
\cite{Kac66,Engman:2005is}, can be paraphrased as:
{\em ``can one hear the stability of a Black Hole horizon?''}.
Although the exact resolution of the spectral problem is a formidable task
in the generic case, semi-classical 
tools (e.g. \cite{Berry81}) may offer relevant insight into the statistical
properties of the spectrum.

\subsection{Inner and outer pressures and the cosmological constant}
The Young-Laplace law says nothing about the absolute values of 
$p_{\mathrm{inn}}$ and $p_{\mathrm{out}}$~\cite{Note4}.
One can, however, speculate about
the implications of the following two 
possibilities: \\
{\em i)} {\em ``Bubble in a room''}: fix $p_{\mathrm{out}}$
to the pressure existing in the absence of the BH, 
namely the cosmological pressure. Then
\bea
\label{e:bubble_room}
\!\!\!\!\!\!\!\!\!  p_{\mathrm{out}} = p_{\mathrm{cosm}} =  -\Lambda/(8\pi) \ \ , \ \
p_{\mathrm{inn}} = (\lambda_o  -\Lambda)/(8\pi) \ .
\eea   
{\ ii)} {\em ``Casimir-like effect''}: Eqs. (\ref{e:SpectralShift})
and (\ref{e:pressure_diff}) imply $p_{\mathrm{inn}}-p_{\mathrm{out}} = 
-\Lambda/(8\pi) - (-\lambda_o^*)/(8\pi)$, motivating the identification
\bea
\label{e:Casimir}
\!\!\!\!\!\!\!\!\!  
p_{\mathrm{inn}} = p_{\mathrm{cosm}}=-\Lambda/(8\pi)  \ \ , \ \
p_{\mathrm{out}} = -\lambda_o^*/(8\pi) \ .
\eea
The outer pressure $p_{\mathrm{out}} = -\lambda_o^*/(8\pi) 
= -(\Lambda + \lambda_o)/(8\pi)$ decreases in the formation
of a stable BH horizon ($\lambda_o\geq 0$). Equivalently, 
an effective (bulk) cosmological constant 
$\Lambda_{\mathrm{eff}}\equiv\Lambda + \lambda_o$  increases 
due to the presence of an inner BH boundary. 
This provides an ingredient
for a physical mechanism correlating the increase of the (effective)
cosmological constant to BH  cosmological dynamics
(note the similarities of such $\Lambda_{\mathrm{eff}}$-''enhancing''
mechanism with the ``neutralization'' of 
$\Lambda$ through the quantum creation of closed membranes
\cite{Brown:1987dd}).

\subsection{BH volume}
A thermodynamic notion of BH volume has been formulated
\cite{Kastor:2009wy,Cvetic:2010jb} by considering
the cosmological constant as an independent intensive variable 
in the BH first law, so that a volume $V$ is introduced as its
corresponding conjugate extensive variable. The present Young-Laplace 
fluid analogy suggests to ``shift'' $-\Lambda/(8\pi)$ to
$\lambda_o/(8\pi)=\Delta p$ [cf. Eq. (\ref{e:SpectralShift})], as 
the appropriate intensive variable
to be employed. That is
\bea
\label{e:BH_volume}
\delta M = T \delta S + 
\Omega_i \delta J_i + \Phi_\alpha \delta Q_\alpha + V_{_{\mathrm{BH}}} 
\delta (\lambda_o/(8\pi)) \ ,
\eea
where $M$ corresponds to a BH {\em enthalpy} and $V_{_{\mathrm{BH}}}$
is now a volume explicitly associated with the BH.
Interestingly, as in \cite{Cvetic:2010jb}, such a thermodynamic
volume provides the Euclidean $V_{_{\mathrm{BH}}}=4\pi/3 \cdot r_{_{\mathrm{H}}}^3$
in ($3$-dimensional) spherical symmetry.

\subsection{BH rest-frame}
The BH Young-Laplace “law” holds for a preferred (local) 
spacetime slicing $\{\Sigma_t\}$. This suggests the proposal: \\
{\em i)} {\em A ``BH rest-frame'' is introduced as the one in which
$H = -\theta^{(k)}$ is constant and the BH Young-Laplace “law” holds}. \\
{\em ii)} {\em Given a unit vector $\xi^a$ tangent to the preferred 
3-slice $\Sigma_t$, but transverse (i.e. admitting normal components) to the horizon 
section ${\cal S}_t$, 
a quasi-local linear momentum along $\xi^a$ is proposed as the dipolar 
part of the mean curvature $H$}
\bea
\label{e:quasilocal-linear-momentum}
P(\xi) 
\equiv\frac{1}{8\pi}\int_{{\cal S}_t} (\xi^a s_a) H  dA \ .
\eea

A horizon slicing is fixed by setting
the value of $D^a \Omega^{(\ell)}_a$. In \cite{AshBeeLew02}
a ``natural'' BH rest-frame was introduced by choosing a vanishing divergence. 
From point {\em iii)} in Theorem 1, 
the present Young-Laplace proposal amounts to a geometric choice
in terms of the principal eigenfunction: 
$D^a \Omega^{(\ell)}_a = D^aD_a\ln\sqrt{\phi_o}$.
Finally note that $P(\xi)$, devised for measuring a 
vanishing linear momentum in the BH rest-frame,
is just the dipolar part of the Brown-York quasi-local energy  \cite{BroYor93}.


\medskip

\smallskip\noindent\emph{Acknowledgments.~} 
I thank M. Reiris for sharing his result on $\lambda_o$ 
and M. Mars for key insights. I thank A. Ashtekar, 
R. Emparan, A. Harte, B. Krishnan, 
F. Pannarale, I. R\'acz, L. Rezzolla and W. Simon for discussions.





\end{document}